\documentclass{svproc}
\usepackage{url}

\begin{document}
\mainmatter              
\title{Dressing for a vector modified KdV hierarchy}
\titlerunning{Dressing for vmKdV hierarchy}  
%
\author{Panagiota Adamopoulou\inst{1} \and Georgios Papamikos\inst{2}}
\authorrunning{P. Adamopoulou et al.} 
%
\tocauthor{Panagiota Adamopoulou and Georgios Papamikos}
\institute{Heriot-Watt University, Edinburgh EH14 4AS, UK,\\
\email{p.adamopoulou@hw.ac.uk},\\ 
\and
University of Essex, Colchester CO4 3SQ, UK, \\
\email{g.papamikos@essex.ac.uk} 
}

\maketitle              

\begin{abstract}
We present the hierarchy and soliton solutions associated to a multi-component generalisation of the modified Korteweg-de Vries equation. A recursive formula for obtaining the Lax operators associated to the higher flows of the hierarchy is provided. Using the method of rational dressing and the symmetries of the Lax operators we obtain the one-soliton solution. We also derive the general rank one-breather solution and express it in terms of determinants. We present the simplest case of the one-breather solution, which is parametrised by two constant unit vectors normal to each other. Finally, we obtain the associated B{\"a}cklund transformation for the hierarchy.

\end{abstract}

\section{Introduction}
Since the establishment of the modern theory of integrable systems there have appeared many generalisations of soliton equations in both independent and dependent variables, such as extensions in $2+1$ variables, systems of equations, and equations with non-commutative variables. Multi-component integrable equations and their hierarchies have also attracted much attention due to their rich mathematical structure and appearance in applications. In the current paper we are concerned with the hierarchy associated to the following multi-component generalisation of the modified KdV equation 
\begin{equation} \label{vmkdv}
\mathbf{u}_{t} + \mathbf{u}_{xxx} + \frac{3}{2} \| \mathbf{u}\| ^2 \mathbf{u}_x = 0 \,, \quad \mbox{with} \quad \mathbf{u} = \mathbf{u}(x,t) \in \bbbr^N \,, 
\end{equation}
and its soliton solutions obtained via the Darboux-dressing scheme. In (\ref{vmkdv}), $\|{\bf u} \| $ denotes the standard Euclidean norm in $\bbbr^N$ and the subscripts denote partial differentiation with respect to the corresponding variables. For convenience, in what follows we will denote by ${\bf u}_j$ the $j$th partial derivative with respect to $x$, with ${\bf u}_0 = {\bf u}$. We remark that equation (\ref{vmkdv}), which has also appeared in \cite{Sokolov Wolf}, is not the vector mKdV equation associated to the vector NLS hierarchy. Several other multi-component generalisations of the mKdV equation have also been studied in e.g. \cite{Fo-Ath,Iwao-Hirota}. In \cite{GG} the authors derived the hierarchy of commuting flows associated to equation (\ref{vmkdv}) based on the Drinfel'd-Sokolov scheme \cite{DS}, and further presented the soliton solution as well as the breather solutions of general rank for the hierarchy.  Here, we additionally present the B{\"a}cklund transformation for the hierarchy, as well as an explicit formula for the simplest one-breather solution.

\section{Higher vector mKdV flows}
Equation (\ref{vmkdv}) admits an infinite number of conservation laws, as shown in \cite{GG} and also discussed in \cite{gptp}. Additionally, the vmKdV is invariant under the Lie symmetries 
\begin{eqnarray}
\widetilde{x} &=& x+ \alpha \,, \nonumber \\
\widetilde{t} &=& t + \beta \,, \nonumber \\
(\widetilde{x}, \widetilde{t}, \widetilde{{\bf u}}) &=& (e^{\epsilon}x, e^{3 \epsilon} t, e^{- \epsilon}{\bf u}) \,, \nonumber
\end{eqnarray}
and is also $O_N$-invariant, i.e. invariant under $\widetilde{{\bf u}} = A {\bf u}$ with $A \in O_N$. A hierarchy of commuting generalised symmetries can be recursively constructed ${\bf u}_{t_{2n+1}} = \mathcal{R} {\bf u}_{t_{2n-1}}$, $n=1,2,\ldots$, using the recursion operator $\mathcal{R}$, given by \cite{GG,Anco vmkdv,JP vmkdv}
\begin{equation} \label{R}
\mathcal{R} {\bf f} = - D_x^2 {\bf f} - \| {\bf u} \|^ 2 {\bf f} - {\bf u}_1 D_x^{-1} \left( {\bf u}^T {\bf f} \right) -  D_x^{-1} ({\bf u}_1 {\bf f}^T - {\bf f} {\bf u}_1^T) {\bf u}\,, 
\end{equation}
starting from ${\bf u}_{t_1} = {\bf u}_1$. For example, we obtain the vmKdV equation (\ref{vmkdv}) by acting with $\mathcal{R}$ on ${\bf u}_{t_1} = {\bf u}_1$
$$
{\bf u}_{t_3} = \mathcal{R} {\bf u}_{t_1} = - {\bf u}_{3} - \frac{3}{2} \| {\bf u} \|^2 {\bf u}_1 \,,
$$
and similarly from ${\bf u}_{t_5} = \mathcal{R} {\bf u}_{t_3}$ we obtain
$$
{\bf u}_{t_5} =  {\bf u}_{5}  + \frac{5}{2} \| {\bf u} \|^2 {\bf u}_{3} + \frac{5}{2} \| {\bf u}_1 \|^2 {\bf u}_{1} + 5 \left(  {\bf u}^T{\bf u}_1 \right) {\bf u}_{2} + 5 \left(  {\bf u}^T{\bf u}_{2} \right) {\bf u}_{1}  + \frac{15}{8} \| {\bf u} \|^4 {\bf u}_{1} \,. \nonumber
$$

All members of the hierarchy ${\bf u}_{t_{2n+1}} = \mathcal{R}^n {\bf u}_1$  have a zero curvature representation $[ \mathcal{L}(\lambda), \mathcal{A}_{2n+1}(\lambda)]=0$, where the linear operators $\mathcal{L}(\lambda)$ and $\mathcal{A}_{2n+1}(\lambda)$ are of the form
\begin{equation} \label{Lax ops}
\mathcal{L}(\lambda) = D_x -  \mathcal{U}(\lambda) \,, \quad  \mathcal{A}_{2n+1}(\lambda) = D_{t_{2n+1}} - \mathcal{V}_{2n+1}(\lambda)\,, \quad n = 1, 2, \ldots\,.
\end{equation}
Here $\lambda$ is a spectral parameter, $D_x$ and $D_{t_{2n+1}}$ denote differentiation with respect to $x$ and $t_{2n+1}$, respectively, and $\mathcal{U}(\lambda)$ and $\mathcal{V}_{2n+1}(\lambda)$ take values in $so_{N+2}(\mathcal{D})[\lambda]$, where $\mathcal{D}$ is the differential ring $\bbbr[{\bf u}, {\bf u}_1, \ldots]$, with $D_x {\bf u}_j = {\bf u}_{j+1}$. The Lax operators (\ref{Lax ops}) are invariant under a $\bbbz_2 \times \bbbz_2 \times \bbbz_2$  reduction group (see e.g. \cite{Mikh RG2}), and in particular they satisfy 
\begin{equation} \label{red grp}
\mathcal{X}(\lambda)^{\dagger} = - \mathcal{X}(\lambda)\,, \quad \mathcal{X}(\lambda^*)^* = \mathcal{X}(\lambda)\,, \quad Q \mathcal{X}(-\lambda)Q^{-1} = \mathcal{X}(\lambda)\,,
\end{equation}
where $^{\dagger}$ denotes the formal adjoint operator, $^*$ denotes complex conjugation, and $Q = {\rm diag}(-1,1, \ldots, 1)$. The matrix $\mathcal{U}(\lambda)$ in (\ref{Lax ops}) is of the form $\mathcal{U}(\lambda) = \lambda J + U$ with \cite{JP vmkdv}
\begin{equation} \label{J component}
J = \left( {\begin{array}{ccc}
0 & 1 & {\bf 0}^T \\
-1 & 0 &  {\bf 0}^T \\
 {\bf 0} & \bf 0 & 0_N\\
\end{array}} \right ),
\quad
U= \left( {\begin{array}{ccc}
0 & 0 & {\bf 0}^T \\
0 & 0 & {\bf u}^T \\
{\bf 0} & - {\bf u} & 0_N  \\
\end{array}} \right ),
\end{equation}
where ${\bf 0}$ is the $N$-dimensional zero vector and $0_N$ the $N \times N$ zero matrix. The $\mathcal{V}_{2n+1}(\lambda)$ are found recursively starting from $\mathcal{V}_1(\lambda) = \mathcal{U}(\lambda)$ according to
\begin{equation} \label{Vk}
\mathcal{V}_{2 n+1}(\lambda) = \lambda^2 \mathcal{V}_{2 n-1} + \lambda A_{2n-1} + B_{2n-1}\,, \quad n=1,2, \ldots \,,
\end{equation}
where 
\begin{eqnarray}
A_{2n-1} &=& - [J , U_{t_{2n-1}} ] - D_x^{-1}({\bf u}^T {\bf u}_{t_{2n-1}}) J \,, \nonumber \\
B_{2n-1} &=& -D_x U_{t_{2n-1}} - D_x^{-1}({\bf u}^T {\bf u}_{t_{2n-1}}) U + D_x^{-1} \left[ D_x U_{t_{2n-1}}, U \right]  \,.\nonumber
\end{eqnarray}
For example, we have that
\begin{equation} \label{V3}
\mathcal{V}_3(\lambda) = \lambda^2 \mathcal{U}(\lambda) - \lambda \left([J, D_x U] + \frac{ \| {\bf u}\| ^2}{2} J   \right) - D_x^2 U - \frac{ \| {\bf u}\| ^2}{2} U + [D_x U,U] \,,
\end{equation}
such that the compatibility condition $[\mathcal{L}(\lambda), \mathcal{A}_{3}(\lambda)]=0$ is equivalent to vmKdV equation (\ref{vmkdv}).

\section{Soliton solutions for the vector mKdV hierarchy}
Employing the method of rational dressing (see for example \cite{vSG dress,Zakharov Shabat II}), we construct soliton and breather solutions to the vmKdV hierarchy. Given that $\Psi(x,t_{2n+1},\lambda)$ is the fundamental solution to the overdetermined compatible system of linear equations
\begin{equation}\label{linear sys}
\mathcal{L}(\lambda) \Psi= 0 \,, \quad \mathcal{A}_{2n+1}(\lambda) \Psi =0 \,, \quad  n=1,2,\ldots \,,
\end{equation}
then a Darboux transformation
$$
\Psi \rightarrow \Phi = M(\lambda) \Psi \,, \quad \textrm{with} \quad \det M(\lambda) \neq 0\,,
$$ 
is a gauge transformation that leaves the linear equations form invariant 
\begin{equation}\label{linear sys 2}
\widetilde{\mathcal{L}}(\lambda) \Phi = 0 \,, \quad \widetilde{\mathcal{A}}_{2n+1}(\lambda) \Phi =0 \,,
\end{equation}
where $\widetilde{\mathcal{L}}(\lambda)  =  D_x - \widetilde{\mathcal{U}}(\lambda)$ with $\widetilde{\mathcal{U}}(\lambda) = \mathcal{U}(\widetilde{\mathbf{u}}, \lambda)$, and similarly for $\widetilde{\mathcal{A}}_{2n+1}(\lambda)$. Here $\widetilde{\mathbf{u}}$ is a new exact solution of the vmKdV hierarchy. The Darboux matrix $M(\lambda)$ can be a constant matrix with respect to the independent variables and it will result to the $O_N$-invariance of the vmKdV hierarchy, or it can be a non-constant matrix which leads to Darboux-dressing relations or to a dressing chain (B{\"a}cklund transformation). From the compatibility condition of (\ref{linear sys}) and (\ref{linear sys 2}) it follows that the Darboux matrix $M(\lambda)$ satisfies the Lax-Darboux equations
\begin{equation} \label{BT rels}
D_x M = \widetilde{\mathcal{U}}M - M\mathcal{U}\,, \quad D_{t_{2n+1}} M = \widetilde{\mathcal{V}}_{2n+1}M - M\mathcal{V}_{2n+1} \,,
\end{equation}
or equivalently, 
\begin{equation} \label{Lax - D}
\widetilde{\mathcal{L}}= M \mathcal{L}M^{-1} \,, \quad \widetilde{\mathcal{A}}_{2n+1} = M \mathcal{A}_{2n+1} M^{-1} \,.
\end{equation}
We further require that the Darboux matrix also satisfies the relations
\begin{equation}  \label{M prop} 
M(\lambda) M(\lambda)^{T} = 1\,,  \quad M(\lambda) = M(\lambda^*)^{*}  \,, \quad  M(\lambda)  =  Q M(-\lambda) Q^{-1} \,,
\end{equation}
which follow from the symmetries (\ref{red grp}) of the Lax operators. The Darboux matrices that lead to the one-soliton and one-breather solutions of the vmKdV hierarchy are parametrised by a complex number and an element of the complex Grassmannian $Gr(s,\bbbc^{N+2})$, with $s=1, 2, \ldots, N+1$ in the case of a breather solution, and $s=1$ for a soliton solution. Further details are presented in \cite{GG}, see also \cite{vSG dress} in relation to the dressing for the vector sine-Gordon equation.

\subsection{Soliton solution: Dressing and B{\"a}cklund transformations}
We consider a Darboux matrix with rational dependence on the spectral parameter $\lambda$,  satisfying relations (\ref{M prop}).  The soliton solution corresponds to a Darboux matrix with two simple poles at $\lambda = \pm i \mu$ of the form \cite{GG,vSG dress,vSG DT}
\begin{equation} \label{DT soliton}
M(\lambda) = 1 + \frac{2 i \mu}{\lambda - i \mu} P -  \frac{2 i \mu}{\lambda +i  \mu}Q P Q \,, \quad {\rm with} \quad P^{*} = Q P Q \,,
\end{equation}
where the matrix $P$ is a projector of ${\rm rank}(P)=1$ of the form
\begin{equation} \label{P soliton}
P = \frac{Q \mathbf{q} \mathbf{q}^T}{\mathbf{q}^T Q \mathbf{q}} \quad {\rm with} \quad \mathbf{q}^T \mathbf{q} = 0 \,, 
\end{equation}
and $\mathbf{q}(x,t_{2n+1}) \in {\bf CP}^{N+1}$. From equations (\ref{Lax - D}) we obtain explicit expressions for ${\bf q}$ in terms of the fundamental solution of the linear systems (\ref{linear sys}), as well as the dressing transformation ${\bf u} \rightarrow  \widetilde{\bf u}$ that leads to soliton solutions for the vmKdV hierarchy. In particular, from the constant in $\lambda$ terms of the first equation in (\ref{Lax - D}) and the double pole at $\lambda = i \mu$ of both equations in (\ref{Lax - D}) we obtain respectively
\begin{equation} \label{L-D new}
\widetilde{U} = U - 2 i \mu \, [ J, P - P^*] \,, \quad P \mathcal{L}(i \mu) P^T = 0 \,, \quad P \mathcal{A}_{2n+1}(i \mu) P^T = 0 \,.
\end{equation}

Using the first equation in (\ref{L-D new})  to express $\widetilde{U}$ in terms of $U$ and $P$ leads to the dressing transformation for the hierarchy
\begin{equation} \label{dressing sol}
\widetilde{u}_j = u_j - 4 i \mu \,  \frac{q_1 \,q_{j+2}}{-q_1^2 + \sum_{k=2}^{N+2} q_k^2} \,, \quad j = 1,2, \ldots, N \,,
\end{equation}
with $q_j$ the components of ${\bf q}$, while the other two equations provide conditions for ${\bf q}$. More precisely, we obtain that the complex vector $\bf{q}$ satisfies 
\begin{equation} \label{Lax q}
\mathcal{L}(i \mu) {\bf q} = 0, \quad \mathcal{A}_{2n+1} (i \mu) {\bf q} =0 \,,
\end{equation}
which imply that 
\begin{equation} \label{q final sol}
{\bf q} = \Psi(i \mu) {\bf C} \,, \quad \rm{with} \quad {\bf C}^T {\bf C} = 0 \,,
\end{equation}
where $\Psi(i \mu)$ is the fundamental solution of the linear systems (\ref{linear sys}) at $\lambda = i \mu$ and ${\bf C} \in \bbbc^{N+2}$ constant.

\noindent \textbf{Example: One-soliton solution}
Starting from the trivial solution ${\bf u} ={\bf 0}$ the fundamental solution of  (\ref{linear sys}) is given by 
$$
\Psi = \exp (\xi J) \,, \quad \textrm{with} \quad \xi(t_{2n+1}, \mu)= \sum_{n=0}^{\infty} (-1)^n \mu^{2 n+1} t_{2 n +1}\,,
$$
with $t_1=x$. The solution $\Psi$ takes the form
$$
\Psi=  
\left( {\begin{array}{ccc}
\cosh \xi & i \sinh \xi & 0^T \\
-i \sinh \xi & \cosh \xi & 0^T \\
0 &  0 & 1_N \\ 
 \end{array} } \right)\,.
$$
Further, the conditions $P^* = QPQ$ and ${\bf C}^T {\bf C} = 0$ in (\ref{DT soliton}) and (\ref{q final sol})  imply that ${\bf C} = (i, c_0, {\bf c}^T)^T$, with $c_0 \in \bbbr$, and ${\bf c} \in \bbbr^N$ a constant vector such that $c_0^2+ \| {\bf c} \|^2 = 1$, see \cite{GG,vSG DT}. Thus, the vector ${\bf q}$ takes the form
\begin{equation} \label{q Psi}
{\bf q}= (i \cosh \xi+ i c_0 \sinh\xi, c_0 \cosh\xi + \sinh\xi, {\bf c}^T )^T \,.
\end{equation}
Then, the dressing formula (\ref{dressing sol}) leads to the one-soliton solution for the vmKdV hierarchy
$$
{\widetilde {\bf u}} = \frac{2 \mu \,  {\bf c}}{\cosh \xi+  c_0 \sinh\xi} \,.
$$

We can also use the Darboux matrix (\ref{DT soliton}) in order to derive the B{\"a}cklund transformation for the vmKdV hierarchy. To this end, we first express ${\bf q}$ in the form
\begin{equation}
{\bf q} = (i, a_0 , {\bf a}^T)^T \,, \quad \mbox{with} \quad a_0^2 + \| {\bf a} \|^2 =1 \,,
\end{equation} 
and $a_0 \in \bbbr$, ${\bf a} \in \bbbr^N$. Using the first equation in (\ref{L-D new}) to express the components of matrix $P$ in terms of the components of $U$ and $\widetilde{U}$ and, given the form of $P$ in (\ref{P soliton}), we obtain the following relation
\begin{equation}\label{BT eq1}
{\bf a} = \frac{{\bf \widetilde{u}} - {\bf u}}{2 \mu} \,.
\end{equation}
Then, from the residue of the simple pole at $\lambda = i \mu$ of the Lax-Darboux equations (\ref{BT rels}) we find that ${\bf a}_x = - \mu a_0 {\bf a} - a_0 {\bf u}$, and thus combining with (\ref{BT eq1}) we obtain
\begin{equation} \label{BT eq2}
(\widetilde{{\bf u}} - {\bf u})_x = - \mu a_0 (\widetilde{{\bf u}} + {\bf u}) \,, \quad a_0^2 + \frac{1}{4 \mu^2}\| \widetilde{{\bf u}} - {\bf u} \|^2  =1 \,.
\end{equation}

\subsection{Breather solution}
The breather solution corresponds to a Darboux matrix $M(\lambda)$ with four poles at $\lambda = \pm \mu, \pm \mu^*$, with $\mu$ a generic complex number, of the form \cite{GG,vSG dress}
\begin{equation} \label{M br}
M(\lambda) = 1 + \frac{M_0}{\lambda - \mu} - \frac{QM_0Q}{\lambda + \mu} + \frac{M^*_0}{\lambda - \mu^*} - \frac{QM_0^*Q}{\lambda + \mu^*},
\end{equation}
where 
\begin{equation} \label{{M0 br}}
M_0 =  q^*B q^T + Q  q C q^T + Q q^* D q^T \,, \quad \quad q^T q = 0 \,,
\end{equation}
and $q \in Gr(s,\bbbc^{N+2}) \simeq M_{N+2,s}(\bbbc)/ GL_{s}(\bbbc)$, for $s=1, 2, \ldots, N+1$. Here, $B, C, D \in M_{s,s}(\bbbc)$ are of the form
\begin{equation} \label{bcd}
B = D G^* H ^{* -1}, ~  C = -D^* F^*  H ^{* -1}, ~ D =  - \left(  F H^{-1} F^* + G^* H^{* -1} G^* - H^* \right) ^{-1},
\end{equation}
where the matrices $F, G, H$ are given by
\begin{equation} \label{fgh}
F = \frac{1}{2 \mu}q^T Q q \,, \quad G = \frac{1}{\mu - \mu^*}q^{* T} q  \,, \quad H = \frac{1}{\mu + \mu^*}q^{*T} Q q \,.
\end{equation}

The double pole at $\lambda= \mu$ of the Lax-Darboux equations (\ref{Lax - D}) leads to relations
$$
\mathcal{L}(\mu) q = 0, \quad \mathcal{A}_{2n+1} (\mu) q =0\,, 
$$ 
thus, we can express the matrix $q$ in terms of the fundamental solution of the linear problems $\mathcal{L}(\mu) \Psi =0\,, ~ \mathcal{A}_{2n+1}(\mu) \Psi = 0$ as
\begin{equation}\label{q final}
q = \Psi(\mu) C \,, \quad \rm{with} \quad C^T C = 0 \,,
\end{equation}
and $C \in M_{N+2,s}(\bbbc)$ a constant matrix. The constant in $\lambda$ terms of the first equation in (\ref{Lax - D}) provide the following dressing transformation leading to breather-type solutions of the vmKdV hierarchy 
\begin{equation} \label{dressing br det}
\widetilde{u}_j = u_j - 4 \textrm{Re} \sum_{k,l = 1}^s 
\left| {\begin{array}{ccc}
q_1^k & 0 & 0 \\
0 & q^l_{j+2} & B^*_{kl} - D^*_{kl} \\
0 & q^{* l}_{j+2} & C_{kl}
 \end{array} } \right|
\,, \quad j = 1,2, \ldots, N \,,
\end{equation}
with $B,C,D$ given in (\ref{bcd}), see \cite{GG} for details. Hence, the breather-type solutions for the hierarchy are characterised by the rank $s$ of matrix $q$ as well as the position $\mu$ of the pole of the Darboux matrix $M(\lambda)$ in (\ref{M br}). In the following example we derive the simplest ($s=1$) one-breather solution, and show that it can be expressed as a ratio of determinants and that it is parametrised by two constant unit vectors normal to each other.

\noindent  \textbf{Example: rank one one-breather solution}
In the case $s=1$, the dressing transformation (\ref{dressing br det}) takes the form
\begin{equation} \label{rank 1 br}
\widetilde{u}_j = u_j - 4 \textrm{Re} \frac{\Delta_j}{\Delta}
\,, \quad j = 1,2, \ldots, N \,,
\end{equation}
where $\Delta_j$ and $\Delta$ are the determinants 
$$
\Delta_j = 
\left| {\begin{array}{ccc}
q_1 & 0 & 0 \\
0 & q_{j+2} & H-G \\
0 & q^{*}_{j+2} & F^*
 \end{array} } \right|,
\quad 
\Delta = \left| {\begin{array}{cc}
F & H-G \\
G+H & F^* 
 \end{array} } \right|,
$$
and $F, G, H$ are now scalar quantities. Starting with the trivial solution ${\bf u} = {\bf 0}$, the fundamental solution of the linear system (\ref{linear sys}) at $\lambda = \mu$ takes the form 
$$
\Psi(\mu)  =  
\left( {\begin{array}{ccc}
\cos \xi & \sin \xi & {\bf 0}^T \\
-\sin \xi & \cos \xi & {\bf 0}^T \\
\bf 0 &  {\bf 0} & 1_N \\ 
 \end{array} } \right), \quad \textrm{with} \quad \xi = \sum_{n=0}^{\infty}  \mu^{2 n+1} t_{2 n +1}\,.
$$
The one-breather solution can be written as
\begin{equation} \label{rank 1 br 2}
{\widetilde{\bf u}} = - \frac{4}{\Delta} \textrm{Re} \big( (C_1 \cos \xi + C_2 \sin \xi) (F^* {\bf c} + (G-H) {\bf c}^*)  \big)\,,
\end{equation}
where ${\bf C} = (C_1, C_2, {\bf c}^T)^T$ such that ${\bf C}^T {\bf C} = 0$. The latter condition implies that the real and imaginary parts of vector ${\bf C}$ have the same length, and furthermore they are normal to each other. Using the fact that ${\bf C}$ is in ${\bf C P}^{N+1}$ we can normalise its real and imaginary parts and assume their length is equal to one. It follows that the one breather solution for the vmKdV hierarchy is parametrised by a complex number (the pole of the Darboux matrix \ref{M br}) and an element of the unit tangent bundle $T_1 \left(\bbbs^{N+1} \right)$ of the sphere
$$
T_1 \left(\bbbs^{N+1} \right) = \lbrace ( {\bf v}_1, {\bf v}_2) \in \bbbr^{2(N+2)} | \; \left<{\bf v}_1, {\bf v}_2 \right>=0\,, \;\|{\bf v}_1\| =  \|{\bf v}_2\| =1   \rbrace \,.
$$

For example, in the case where the real and imaginary parts of vector ${\bf C}$ are given by
$$
{\bf C}_R = (1,0, {\bf 0})^T\,, \quad {\bf C}_I = (0, \ldots, 0, 1, 0, \ldots,0)^T\,,
$$
respectively, where  $1$ appears in the $(j+2)$ position in ${\bf C}_I$. Then, starting from the trivial solution ${\bf u}= {\bf 0}$ the breather solution (\ref{rank 1 br 2}) takes the form
\begin{equation}
{\widetilde u}_k = 
\left\{
\begin{array}{lr}
    - \frac{4}{\Delta} \textrm{Re} \big(i \cos \xi (F^* + H -G) \big)\,, \quad k= j ,\\
    0 \,, \quad k \neq j \\
\end{array} 
\right.
\end{equation}
The denominator in the above expression can be written in the form
\begin{equation} \label{denom}
\Delta = - \frac{1}{r^2} \left( \tan \theta \sin^2 A + \frac{\cosh^2 B}{\tan \theta}   \right)^2  ,
\end{equation}
where $r = |\mu|$, $\theta = \arg (\mu)$, $A = \textrm{Re}(\xi)$ and $B= \textrm{Im}(\xi)$. 
Finally, for the particular choice of ${\bf C}$ given above, we obtain the following expression for the rank one one-breather solution and the 
\begin{equation}
\widetilde{u}_j = 4 r \frac{ \sin \theta \sin A \sinh B - \cos \theta \cos A \cosh B}{\tan \theta \sin^2 A + \frac{\cosh^2 B}{\tan \theta}} \,,
\end{equation}
which, as expected, is a breather solution of the scalar mKdV.

\section{Conclusions}
In this paper we presented a vectorial generalisation of the well-known mKdV equation, namely equation (\ref{vmkdv}). It is important to note that this equation is not the third flow of the vector NLS hierarchy, since its Lax pair is characterised by relations (\ref{red grp}), and additionally $J$ in (\ref{J component}) is not a regular element of the underlying Lie algebra, contrary to the Drinfel'd-Sokolov construction \cite{DS}. For equation (\ref{vmkdv}) we presented its recursion operator which can be used to construct all higher flows of the integrable hierarchy, and also a recursive formula that produces the Lax operator associated to each member of the hierarchy. We used the structure of the Lax operators to construct two Darboux transformations and we obtained the one-soliton solution and the general rank one-breather solution. The breather solution, in the simplest case, is a breather solution of the scalar mKdV equation. We leave the study of the more general breather solutions, as well as the $n$-soliton and $n$-breather solution for future work. As part of this goal, we aim to study Hirota's direct method as well as the general scattering problem and its inversion through a Riemann-Hilbert problem.


%
%

\end{document}